\documentclass[aps,prl,twocolumn,groupedaddress,showpacs]{revtex4}
\usepackage{amssymb}
\usepackage{hyperref}
\usepackage{amsmath}
\usepackage{graphicx,color}
\definecolor{r}{rgb}{1,0.33,0.29}
\definecolor{g}{rgb}{0.54,0.76,0.38}
\definecolor{b}{rgb}{0,0,1}

\begin{document}


\title{Divergence of Voronoi cell anisotropy vector: A threshold-free characterization of local structure in amorphous materials} 


\author{Jennifer M. Rieser}
\altaffiliation[Now at ]{School of Physics, Georgia Institute of Technology, Atlanta, GA, 30332-0430, USA}
\affiliation{Department of Physics and Astronomy, University of Pennsylvania, Philadelphia, PA 19104-6396, USA}
\email{jennifer.rieser@gatech.edu}

\author{Carl P. Goodrich}
\altaffiliation[Now at ]{School of Engineering and Applied Sciences, Harvard University, Cambridge, MA 02138, USA}
\affiliation{Department of Physics and Astronomy, University of Pennsylvania, Philadelphia, PA 19104-6396, USA}

\author{Andrea J. Liu}
\affiliation{Department of Physics and Astronomy, University of Pennsylvania, Philadelphia, PA 19104-6396, USA}

\author{Douglas J. Durian}
\affiliation{Department of Physics and Astronomy, University of Pennsylvania, Philadelphia, PA 19104-6396, USA}
\email{djdurian@physics.upenn.edu}


\date{\today}

\begin{abstract}
 
Characterizing structural inhomogeneity is an essential step in understanding the mechanical response of amorphous materials.  We introduce a threshold-free measure based on the field of vectors pointing from the center of each particle to the centroid of the Voronoi cell in which the particle resides.  These vectors tend to point in toward regions of high free volume and away from regions of low free volume, reminiscent of sinks and sources in a vector field.   We compute the local divergence of these vectors, where positive values correspond to \emph{overpacked} regions and negative values identify \emph{underpacked} regions within the material.  Distributions of this divergence are nearly Gaussian with zero mean, allowing for structural characterization using only the moments of the distribution.  We explore how the standard deviation and skewness vary with packing fraction for simulations of bidisperse systems and find a kink in these moments that coincides with the jamming transition.      
\end{abstract}


\maketitle

In glassy liquids and disordered solids, heterogeneities in local structure correlate with heterogeneous particle rearrangement dynamics arising from thermal fluctuations or applied mechanical load~\cite{PatrickRoyall:2008fz, Tsamados:2009ke, Manning:2011dk, Malins:2013br, Jack:2014ba, Cubuk:2015cd}.  Characterizing local structural heterogeneity is therefore important in experiment, for example via the network of contacts and force chains \cite{Majmudar:2005bn, Desmond:2013js, Jorjadze:2013bf, Bassett:2015ks}, and as a step in understanding thermal and mechanical response.  A simple and physically-appealing measure of local structure that forms the basis of historically-important theories of glassy dynamics and plasticity is free volume~\cite{Turnbull:1961jm, Spaepen:1977tc, berthier2011dynamical,Maiti:2014ie}.  Regions that are underpacked have a larger local free volume, and therefore ought to rearrange or yield more easily.  Though intuitive, free volume is inherently a concept based on hard spheres and only applies at densities below jamming.

Here we introduce a generalization of the concept of free volume that derives from the radical Voronoi network and hence applies in a consistent way to particles interacting via any inter-particle potential at any density.  Our measure, $Q_k$, is inspired by the observation that the center of a particle center deviates from the centroid of the corresponding Voronoi cell when the configuration is disordered.  In two dimensions, it is defined as
\begin{align}
Q_k \equiv (\nabla \cdot \mathbf{c}) (A_k / \bar{A}), 
\label{eqn:Qk}
\end{align}
where $\mathbf{c}$ is the interpolated field of vectors that point from particle centers to the corresponding Voronoi cell centroids, the divergence is taken over a Delaunay triangle $k$ with area $A_k$, and $\bar{A}$ is the average of all $A_k$ within the a packing.  By construction, $Q_k$ is dimensionless and has zero mean.  It is sensitive to local structural heterogeneity and -- though purely geometrical -- has a clear physical interpretation: positive/negative values respectively correspond to overpacked/underpacked regions.  In addition to establishing a statistical correlation between $Q_k$ and local relative free volume, we find that the distribution of $Q_k$ values over a packing is nearly Gaussian, with  mode and median nearly equal to the mean (zero); hence it may be well-described by just the standard deviation and the skewness.

As an illustration, we calculate $Q_k$ for a system of soft disks at a series of packing fractions that are widely varied, from the dilute limit to well above the jamming transition.  Morse and Corwin~\cite{Morse:2014hg} have recently identified geometrical features similarly based on Voronoi tessellations that exhibit singularities at the jamming transition.  Here we find the standard deviation and skewness of the $Q_k$-distribution also exhibit kinks at the transition.   Thus there is a signature of the jamming transition in $Q_k$, a geometrical quantity with clear physical relevance.

\begin{figure}[h]
\includegraphics[width=\columnwidth]{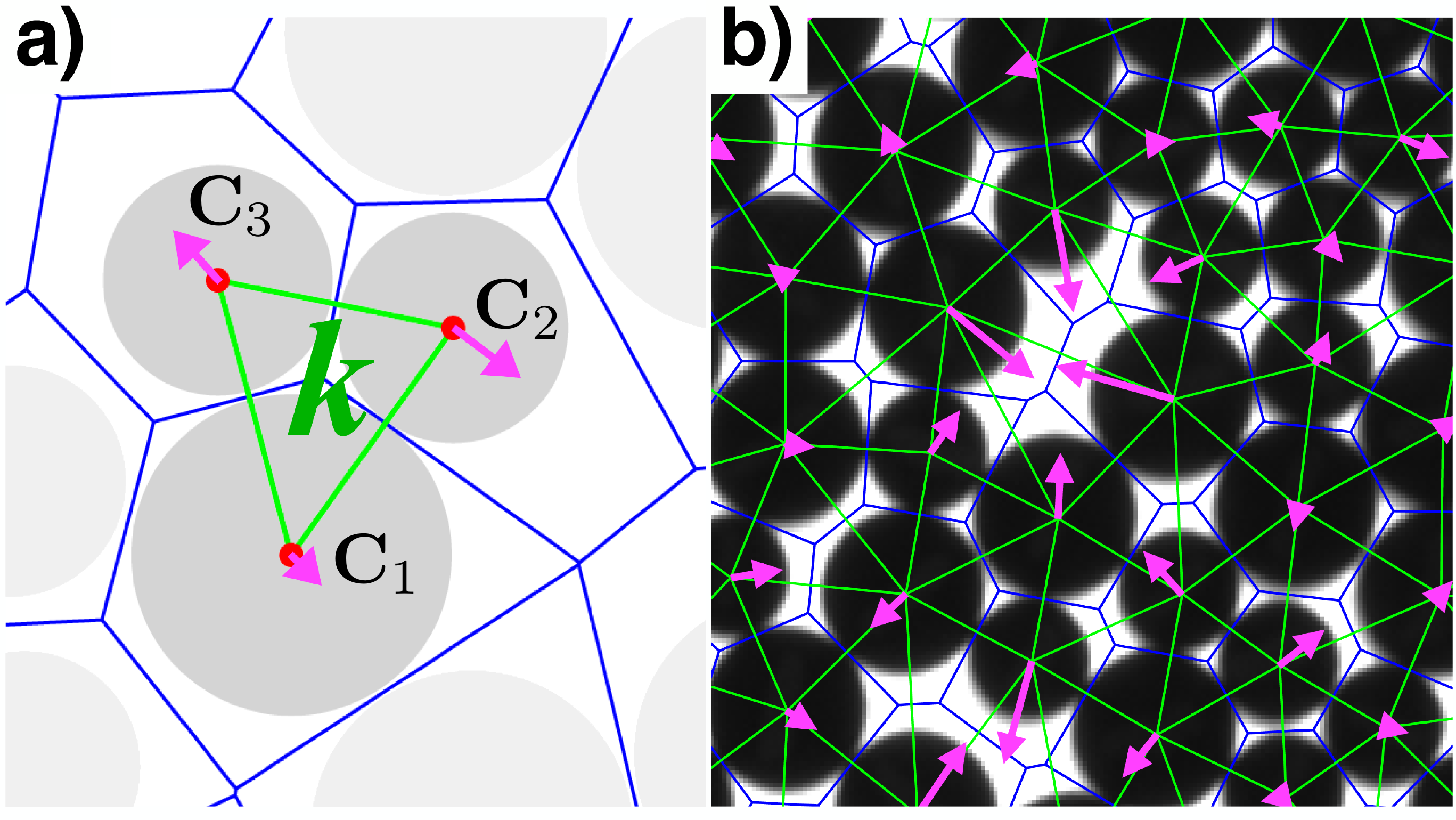}
\caption{(color online) Particle packing from (a) simulation and (b) experiment \cite{Li:2015kt,RieserExpInPrep}, with superimposed radical Voronoi tesselation (blue) and Delaunay triangulation (green).  Also shown are vectors $\mathbf{C}_p$ (magenta) that point from each particle center (red dot) to the centroid of its Voronoi cell.  In (b), these are elongated by a factor of $8$ for each of visualization.}
\label{fig:definitions}
\end{figure}

We begin by using the software package {\tt voro++}~\cite{Rycroft:2009ff} to determine the radical Voronoi tessellation, a space-filling generalization of the standard Voronoi construction to polydisperse systems.  In this framework, each cell edge is determined from two adjacent circles, and is given by the locus of points from which tangent lines drawn to both circles have the same length.    If all particle radii are equal, the standard Voronoi tessellation is recovered.  Fig.~\ref{fig:definitions} shows an example, where the Voronoi tessellation is overlaid on an image of particles.  Any two particles with a shared Voronoi cell edge are defined as neighbors, and from this, we generate a generalized Delaunay triangulation by connecting groups of three mutual neighbors into triangles, as shown in green in Fig.~\ref{fig:definitions}. 

The position of a particle within its Voronoi cell is an indicator of local variation in the packing.  This motivates consideration of a local anisotropy vector, $\mathbf{C}_p$, for each particle $p$, that points from the center of the particle to the centroid of its Voronoi cell, as shown by the magenta arrows in Fig.~\ref{fig:definitions}.  In monodisperse crystalline packings with a single particle per unit cell, each particle center and the corresponding Voronoi cell centroid coincide; therefore $\mathbf{C}_p=0$ for all $p$, consistent with the idea that $\mathbf{C}_p$ is a measure of the structural anisotropy. This vector is one of several Minkowski functionals~\cite{SchroderTurk:2010jc} associated with a Voronoi cell, many of which have been used to describe packing heterogeneity~\cite{SchroderTurk:2010cr,Schaller:2015gq}.  Additionally, it has been discussed in the context of structure in liquids~\cite{Rahman:1966ix,Farago:2014ec},and has been found to be correlated with particle motion \cite{Rahman:1966ix, Slotterback:2008hi}.  As might be expected, $\mathbf{C}_p$ points in the direction of excess free volume, indicating the direction in which the particle has the most space to move.  However, local spatial variations of this vector have not been previously explored.

Fig.~\ref{fig:definitions}b shows a typical example of $\mathbf{C}_p$ vectors for several particles in a bidisperse packing with particle size ratio of $3$:$4$ and hard-sphere interactions.  Vectors tend to point in toward locally less well-packed and away from locally more well-packed regions of the packing, reminiscent of sinks and sources in a vector field.  Therefore it is natural to consider the divergence of a field defined by interpolating the $\mathbf{C}_p$ vectors over a local region.  We choose Delaunay triangles as the local regions over which to perform interpolation and differentiation of the $\mathbf{C}_p$ vectors.  This allows us to use the framework of the constant strain triangle of finite element analysis~\cite{FEM} to find local spatial variations of the $\mathbf{C}_p$ vectors.  In particular, each triangle is treated independently and we assume that the associated $\mathbf{C}_p$ vectors define a vector field $\mathbf{c} = (c_{x},c_{y})$ that varies linearly over each triangle:
\begin{align} 
&c_x(x,y) = d_x+ d_{xx} x + d_{xy} y, \nonumber  \\
&c_y(x,y) = d_y + d_{yx} x + d_{yy} y.
\label{eqn:C}
\end{align}   
For each triangle the six constants $d_x, d_y, d_{xx}, d_{xy}, d_{yx},$  and $d_{yy}$ can be determined by evaluating Eq.~(\ref{eqn:C}) at the triangle vertices and inverting the resulting matrix equation. If the triangle coordinates are shifted so that the centroid of the triangle is located at the origin, the vector $(d_x,d_y)$ is equivalent to $(\bar{c}_x,\bar{c}_y)$, where the averages are taken over the triangle vertices.  The tensor $d_{ij} = \partial c_i /\partial x_j$ is independent of the origin location and describes the spatial variation of the $\mathbf{c}$ field over the triangle, with the divergence given by $\mathrm{Tr}(d_{ij})$.  

\begin{figure}[h]
\includegraphics[width=\columnwidth]{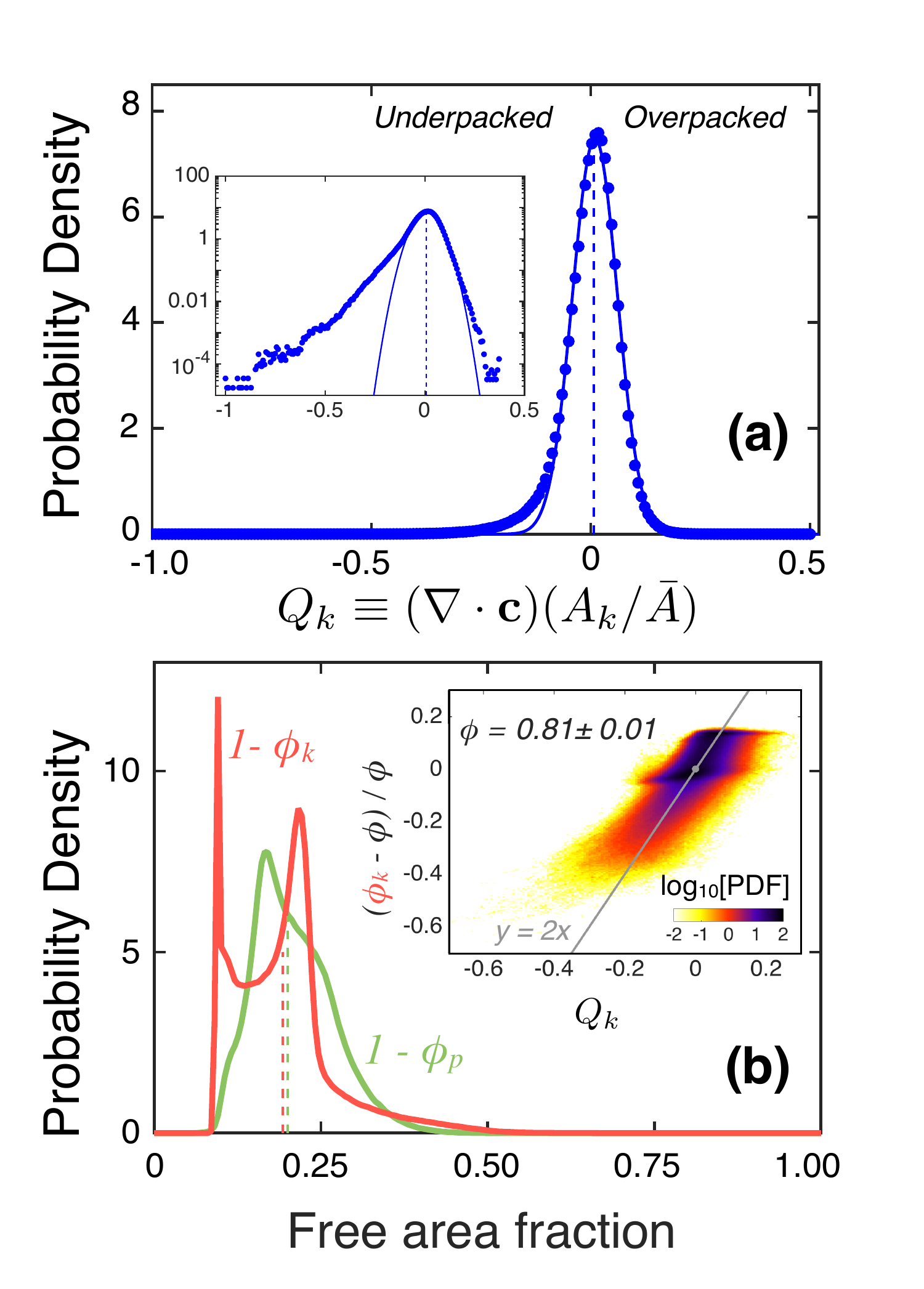}
\caption{(color online) (a) Probability density of normalized divergence of center-to-centroid vectors for the experimental packing of bidisperse hard spheres shown in Fig.~\ref{fig:definitions}b (circles) and best-fit Gaussian (solid curve).  $Q_k > 0$ regions are more tightly packed than their surroundings, hence we call these regions overpacked.   $Q_k < 0$ regions are more loosely packed than their surroundings, and are therefore labeled underpacked.  The inset shows the probability density and Gaussian fit with a logarithmic $y$-axis, highlighting the deviation of the data from Gaussian, particularly in the negative tail.  b) PDFs of triangle-based free area fraction, $1-\phi_{k}$, and Voronoi-based free area fraction, $1-\phi_p$, for the same packing.  These distributions are more complicated and less intuitive than for $Q_k$. Inset: $Q_k$ correlates well with a fractional deviation from the global packing fraction, and therefore is similar to a relative free area.  All dashed lines indicate the median of the data set of the same color.}
\label{fig:pdfs}
\end{figure}

From the divergence theorem, all contributions from interior particles cancel upon performing the sum $\sum_k Q_k$ over all triangles in a packing, leaving only contributions from the boundary particles.  This results in $\langle Q_k \rangle = 0$ for infinite systems and for systems with periodic boundary conditions.  The interpretation of $Q_k$ is then physically intuitive:  triangles with $Q_k < 0$ are less well packed than their surroundings, so we label these regions as \emph{underpacked}, while triangles with $Q_k > 0$ are locally more tightly packed than their surroundings, so we refer to these regions as \emph{overpacked}.  Thus $Q_k$ is a measure of relative free volume, and, as shown below, the statistical correlation is quantitative.

Fig.~\ref{fig:pdfs}a shows the probability density of $Q_k$ for the experimental bidisperse system shown in Fig.~\ref{fig:definitions}b.  The probability density is nearly Gaussian with a small, positive mean resulting entirely from the finite boundaries of the packing.  Thus we may characterize packings to a great degree just from the standard deviation and skewness of the $Q_k$ distribution.  To the extent that the probability density is truly Gaussian and the $Q_k$ values are spatially uncorrelated~\cite{Rieser:SuppMat}, the packing is random in a very simple sense.  But since adjacent triangles share two $\mathbf{C}_p$ vectors, there must be at least short-range correlations.  Nevertheless, $Q_k$ is closer to a Gaussian random variable than any other structural quantity previously used to characterize random packings.  Furthermore, deviations from Gaussianity (e.g.\ underpacked particles in the tail of the distribution) are likely to have important physical consequences \cite{Beuman:2012ca}.

To build intuition, we now compare $Q_k$ with the local area fraction.  Fig.~\ref{fig:pdfs}b shows the probability density for two standard measures, based on particle area per Voronoi cell and per Delaunay triangle.  The probability densities for these quantities have irregular complicated shapes, where the median differs significantly from the mode for each distribution.  There is no clear feature demarcating under- versus over-packed regions.  Nevertheless, the magnitude of $Q_k$ is on the order of -- and statistically correlated with -- the relative free area defined as $(\phi_k-\phi)/\phi$ where $\phi_k$ is the triangle-based area fraction and $\phi$ is the global area fraction of the sample.  This is demonstrated by the contour plot of relative free area versus $Q_k$ in the inset of Fig.~\ref{fig:pdfs}b.  Similar plots in~\cite{Rieser:SuppMat} show that good correlation holds at all packing fractions, but less so for dilute systems due to the development of long tails away from the heart of the distributions.  We may thus consider the size of $Q_k$ as a semi-quantitative indication of local free volume relative to the average packing. 

For the remainder of the paper, we use the $Q_k$ distribution as a tool to characterize structure versus packing fraction for simulated systems in two-dimensions.  Static packings are created using four different protocols.  For the first, a large number $N = 5000$ to $N = 80$,$000$ of points are placed at random in a box.  For each $N$, we find that the average moments of $Q_k$ over $200$ configurations are independent of $N$.  For the second protocol, we generate several packings of non-overlapping monodisperse disks.  Here, each disk has a radius equal to $1$, and a proposed new disk is only accepted and placed in the box if it does not overlap with any existing disks.  These packings contain anywhere from $N = 1000$ to $N = 160$,$000$ disks, spanning area fractions from $\phi \approx  0.003$ to $\phi \approx  0.5$.  At least $40$ packings are generated for each value of $\phi$.  

For the final two protocols, we numerically generate systems composed of $N=2048$ soft repulsive disks with mass $M$. Two particles $i$ and $j$ interact with the pairwise potential
\begin{align}
V_{ij} = \frac{\epsilon}{2} \left( 1 - \frac{r_{ij}}{R_i+R_j} \right)^2 \Theta \left(1 - \frac{r_{ij}}{R_i+R_j} \right),
\end{align}
where $r_{ij}$ is the distance between the particle centers, $R_i$ and $R_j$ are the particle radii, $\Theta(x)$ is the Heaviside step function, and $\epsilon$ sets the energy scale. To prevent crystallization, we use a $50-50$ mixture of particles with a size ratio of $1.4$. The disks are initially placed at random in a periodic simulation box with a packing fraction $\phi$, and are then allowed to move according to two different procedures. The units of length, mass, and energy are $2R_\text{avg}$, $M$, and $\epsilon$ respectively, where $R_\text{avg}\equiv N^{-1}\sum_m R_m$ is the average particle radius. 

For the third protocol, thermalized configurations are generated at a very low temperature using molecular dynamics simulations at constant NVT, performed using LAMMPS~\cite{Plimpton:1995}. Beginning at a temperature of $T_\text{start}=0.05$, we slowly cool the system to $T=10^{-7}$ over $5\times 10^6$ time steps. The system then remains at $T=10^{-7}$ for an additional $10^7$ time steps.   The fourth and final protocol corresponds to an infinitely fast quench from infinite to zero temperature. Beginning with the initial random configuration, we minimize the total energy to a local minimum using the FIRE algorithm~\cite{Bitzek:2006bw}. Each protocol was repeated $500$ times at each packing fraction. 

\begin{figure} [h]
\includegraphics[width=\columnwidth]{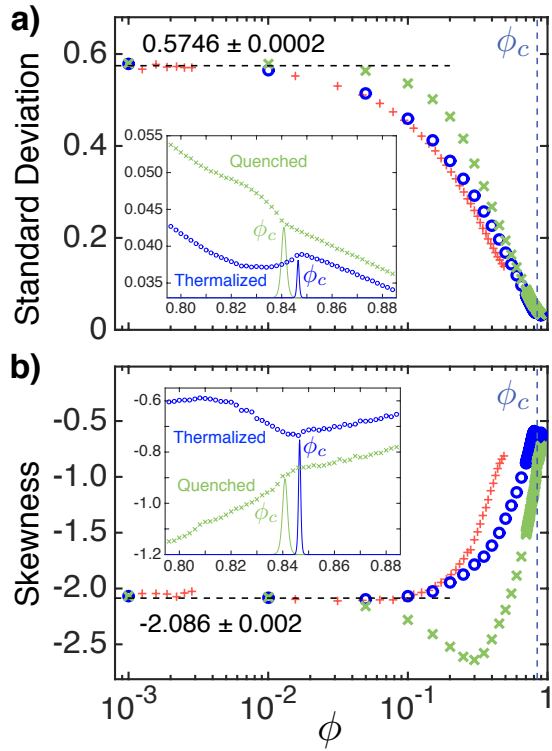}
\caption{(color online) Standard deviation (a) and skewness (b) of the $Q_k$ distributions versus global packing fraction $\phi$.  Each data point represents the average value of the specified moment over several configurations; the associated uncertainty is smaller than the symbol.  The main plots emphasize the low $\phi$ behavior, which approaches the same constant value for all preparation protocols shown: infinitely-fast quench ({\color{g} $\times$}), thermal ({\color{b} $\circ$}), randomly placed non-overlapping monodisperse circles ({\color{r} $+$}), and random point patterns with $\phi=0$ (dashed line).  Insets show behavior near $\phi_c$.  Rescaled distributions of $\phi_c$ are plotted as solid curves, with color indicating preparation protocol from which they were determined.   The moments of $Q_k$ have a kink that coincides with the peak in the respective $\phi_c$ distribution. }
\label{fig:moments}
\end{figure}

For all final configurations, the $Q_k$ distributions and moments are computed.  The low-$\phi$ behavior of the standard deviation is emphasized in the main plot of Fig.~\ref{fig:moments}a.  The results for protocols $2-4$ appear to converge nicely to the value $0.5746\pm0.0002$ found from the random point patterns of protocol $1$.  Fig.~\ref{fig:moments}b shows a similar convergence of the skewness consistent with the value $-2.086\pm0.0002$ obtained for point patterns.  This validates the protocol methods, and serves to establish the low-$\phi$ ``ideal gas'' limiting behavior of the $Q_k$ distributions.  

As $\phi$ increases away from zero, the moments of the $Q_k$ distribution change in a protocol-dependent fashion.  As seen, the thermalized configurations are closer than the rapidly-quenched configurations to the randomly-placed non-overlapping sphere configurations.  The $Q_k$ distributions all become narrower with increasing $\phi$, as shown by the standard deviation plot.  And as judged from the skewness plot, the distributions generally become more Gaussian -- as mentioned earlier.  However the quenched configurations show an initial increase in the skewness magnitude before decreasing towards zero.

The insets in Fig.~\ref{fig:moments} show zoom-ins of the moments near the critical packing fraction $\phi_c$, where the systems become jammed.  For our simulations, there is a distribution of $\phi_c$ values due to the finite system size~\cite{OHern:2002bs,OHern:2003gu}.  Each protocol produces a different distribution~\cite{Chaudhuri:2010jg,Vagberg:2011fy}.  Here, the average and standard deviation are $\phi_c= 0.8409 \pm 0.0012$ for the quenched protocol and $\phi_c= 0.8465 \pm 0.0005$ for the thermalized protocol.  As expected, the $\phi_c$ values are smaller and more widely distributed for quenched configurations, since thermalized configurations have more opportunity to relax~\cite{Chaudhuri:2010jg}.  The unnormalized $\phi_c$ distributions are individually rescaled to reach the respective data curve in the insets of Fig.~\ref{fig:moments}, in order to mark the jamming transitions.

The key striking result evident in the Fig.~\ref{fig:moments} insets is that a signature of the jamming transition exists in the $Q_k$ distributions.  Namely, the standard deviation and the skewness both show a kink where the $\phi_c$ distributions are peaked.  For the quenched protocol, the skewness kink is smallest and may deviate from slightly $\phi_c$.  For the thermalized protocol, the behavior near $\phi_c$ is considerably more dramatic.  In particular, the kinks are extremely pronounced in that the derivatives of standard deviation and of skewness versus $\phi$ actually change sign on opposite sides of the transition.  Furthermore, there is non-monotonic behavior below the transition, with the standard deviation and the skewness exhibiting a minimum and maximum, respectively, at a packing fraction a few percent below $\phi_c$.  As a measure of static structure versus $\phi$, the $Q_k$ distribution is thus even capable of signaling a precursor that the onset of jamming is imminent.   Since this happens for the thermalized but not the quenched configurations, it could be related to the dynamical hard-sphere glass transition; however, the extrema are at different $\phi$ values.  Note that the differences in the trends for the two protocols implies that details of the local structure are sensitive to protocol even though other quantities that are singular at the jamming transition, such as the average contact number, scale the same way with increasing pressure for packings prepared using different protocols~\cite{Chaudhuri:2010jg}.

In conclusion, we have shown that $Q_k$ is a geometrical measure of structure, with the physical meaning of a relative free volume, which displays a strong signature of the jamming transition.  It is particularly easy to interpret $Q_k$ in terms of overpacked and underpacked regions since the $Q_k$-distribution has zero mean, by construction, and is nearly Gaussian for non-dilute systems.  Though all our examples are two-dimensional, the concept of $Q_k$ may be extended to any dimension by appropriate Voronoi construction.  For thermal and sheared systems, there is longstanding interest in identifying structural features that lead to dynamical activity such as heterogeneous particle rearrangements and shear bands.  The correlation of $Q_k$ with dynamics, as well as with structural predictors of rearrangements found by machine learning \cite{Cubuk:2015cd}, may now be studied.  This could give new meaning to the concept of ``free volume," which has been assumed in many theories to affect dynamics and thereby control the glass transition and glassy rheology~\cite{berthier2011dynamical}.

This work is supported by the National Science Foundation through grants MRSEC/DMR-1120901 (JMR, DJD, AJL) and DMR-1305199 (DJD), as well as by a Simons Investigator Award from the Simons Foundation (AJL) and a University of Pennsylvania SAS Dissertation Fellowship (CPG).


\appendix
\renewcommand\thefigure{A\arabic{figure}}
\setcounter{figure}{0}

\begin{figure*} [ht]
     \centering
     \large
    \textbf{Supplemental figures for ``Divergence of Voronoi cell anisotropy vector: A threshold-free characterization of local structure in amorphous materials"}\par\medskip
    \vspace{3mm}
     \includegraphics[width=1.3\columnwidth]{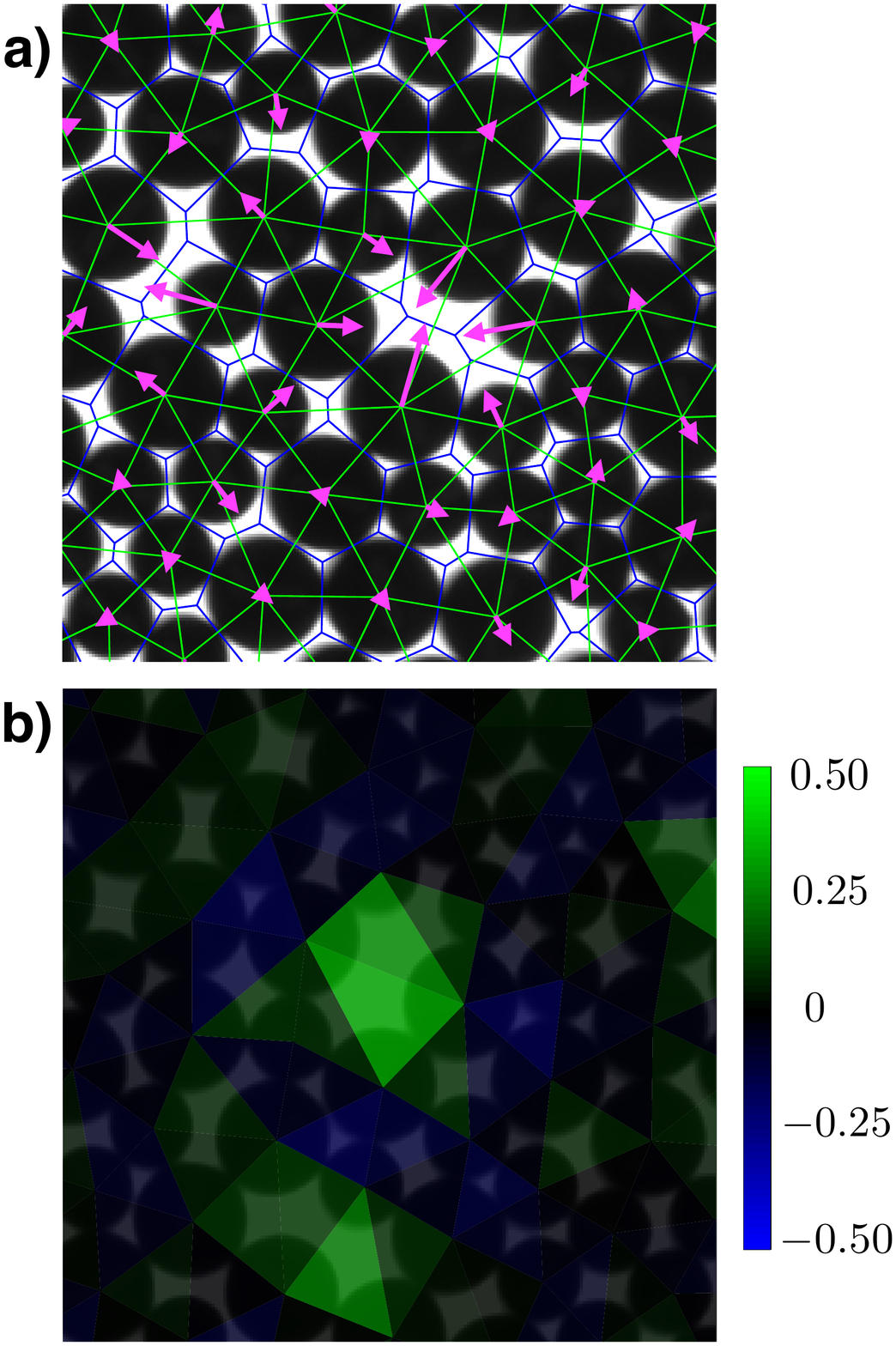}
     \caption{(a) Center-to-centroid vectors $\mathbf{C}$ for an experimental packing, like the one shown in Fig.~1 of the main text, elongated by a factor of 8 for ease of visualization.  (b) Resulting $Q_k$ field for the same region of the same packing.  More loosely packed regions appear green, while locally more well packed regions are blue. }
\end{figure*}


\begin{figure*} [ht]
     \centering
     \textbf{Thermalized}\par\medskip
     \includegraphics[width=1.6\columnwidth]{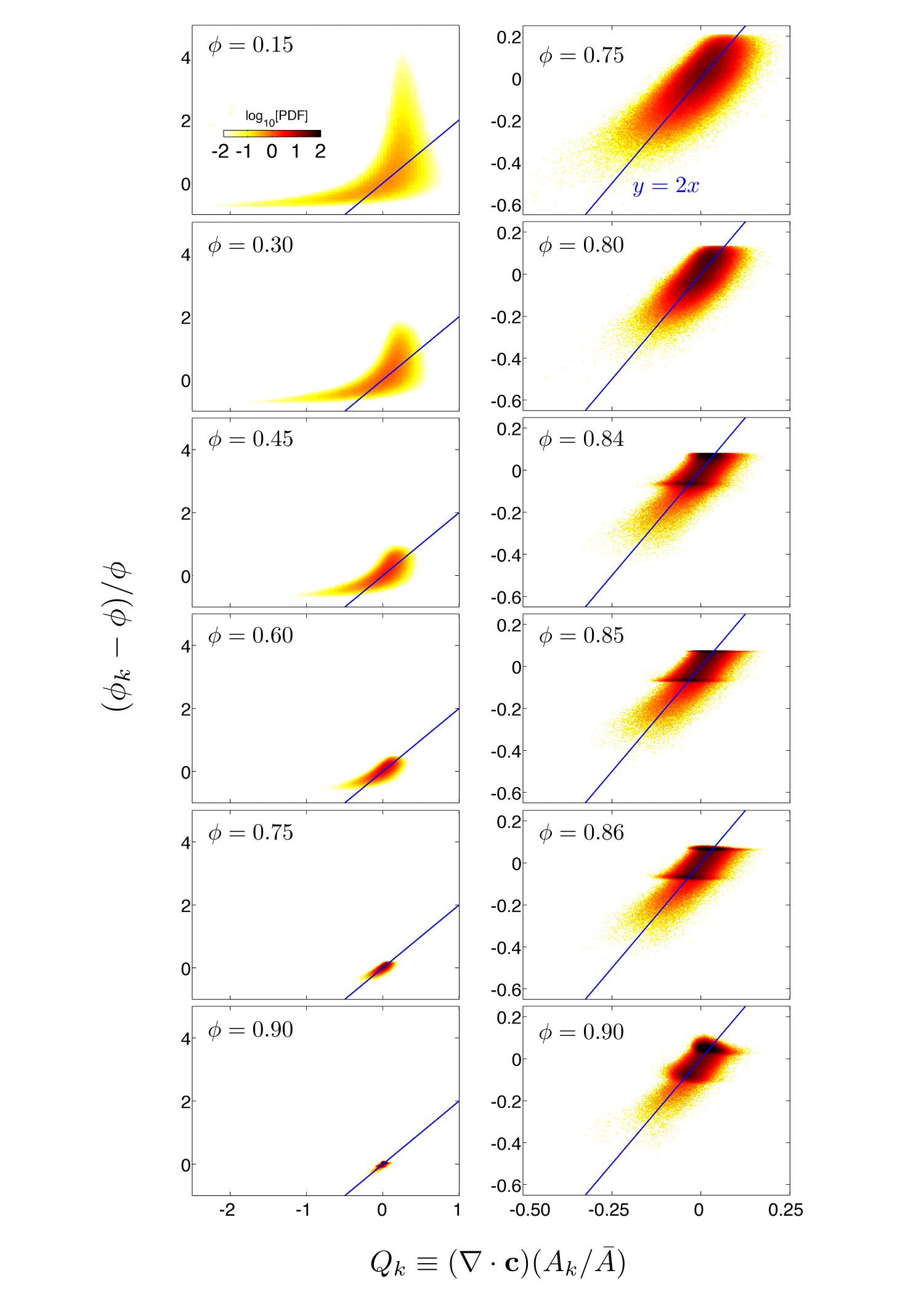}
     \caption{Correlation between the local relative free area and the area-weighted divergence of the center-to-centroid vector field, for numerous global area fractions.  Here, the packings are numerically generated by the thermalization protocol at $T=10^{-7}$.  The blue line is the same in all plots, $y=2x$.  The right column shows a range of packing fractions spanning the jamming transition at $\phi_c=0.8465\pm0.0005$.  Throughout this range the relative free area is highly correlated with $Q_k$, and is twice as large on average.  The left column shows a broader range of $\phi$.  Far below $\phi_c$, the relative free area is still roughly $2Q_k$ on average, but the correlation in the tails becomes progressively weaker.}
\end{figure*}

\begin{figure*} [ht]
     \centering
     \textbf{Quenched}\par\medskip
\includegraphics[width=1.6\columnwidth]{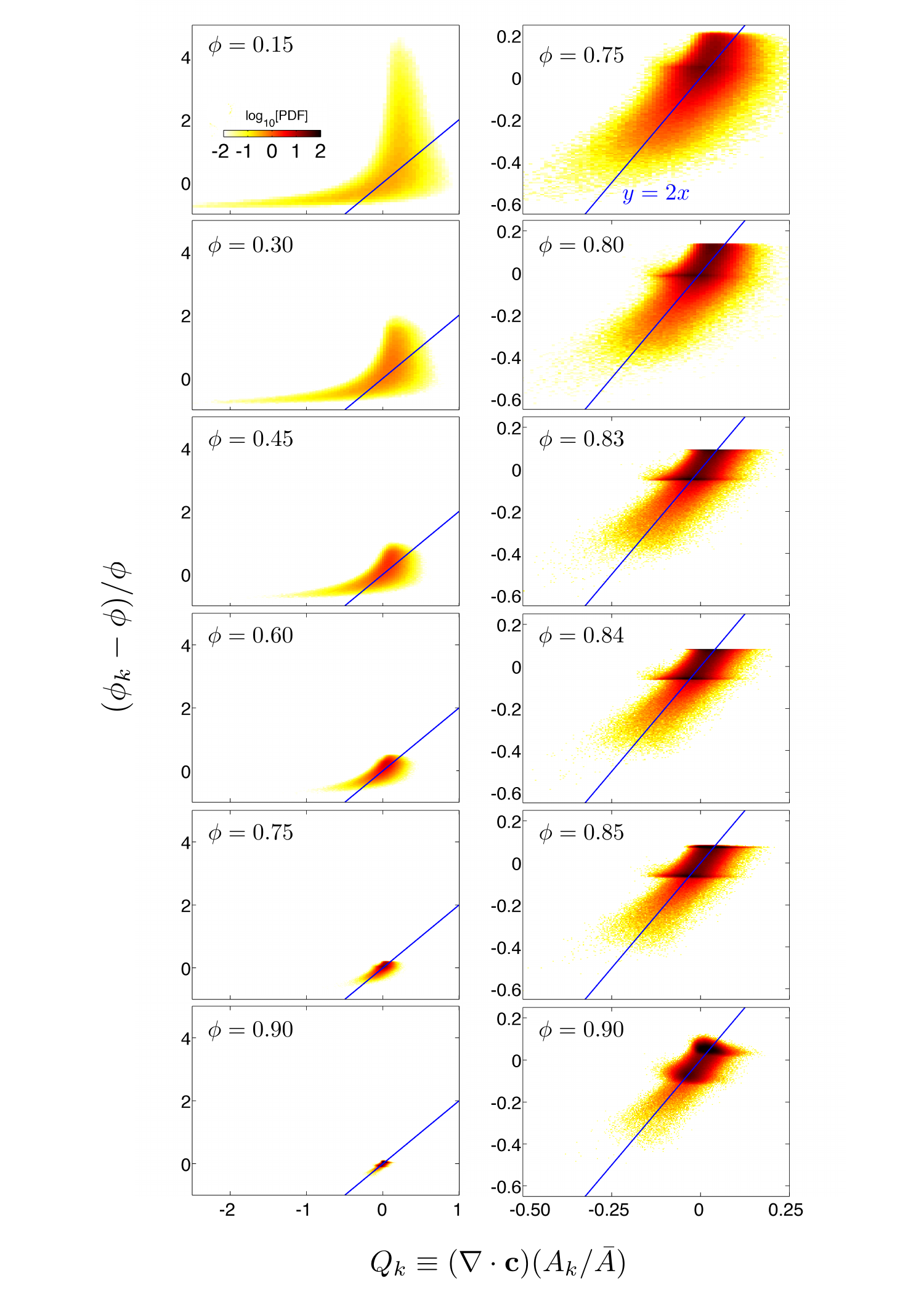}
\caption{Correlation between the local relative free area and the area-weighted divergence of the center-to-centroid vector field, for numerous global area fractions.  Here, the packings are simulated by rapid quench from $T=\infty$ to $T=0$, and the jamming transition is at $\phi_c=0.8409\pm0.0012$.  As seen in the previous plot, the relative free area is about $2Q_k$ on average, at all $\phi$, but with tails that weaken the correlation upon dilution.}
\end{figure*}

\begin{figure*} [ht]
     \centering
     \includegraphics[width=1.6\columnwidth]{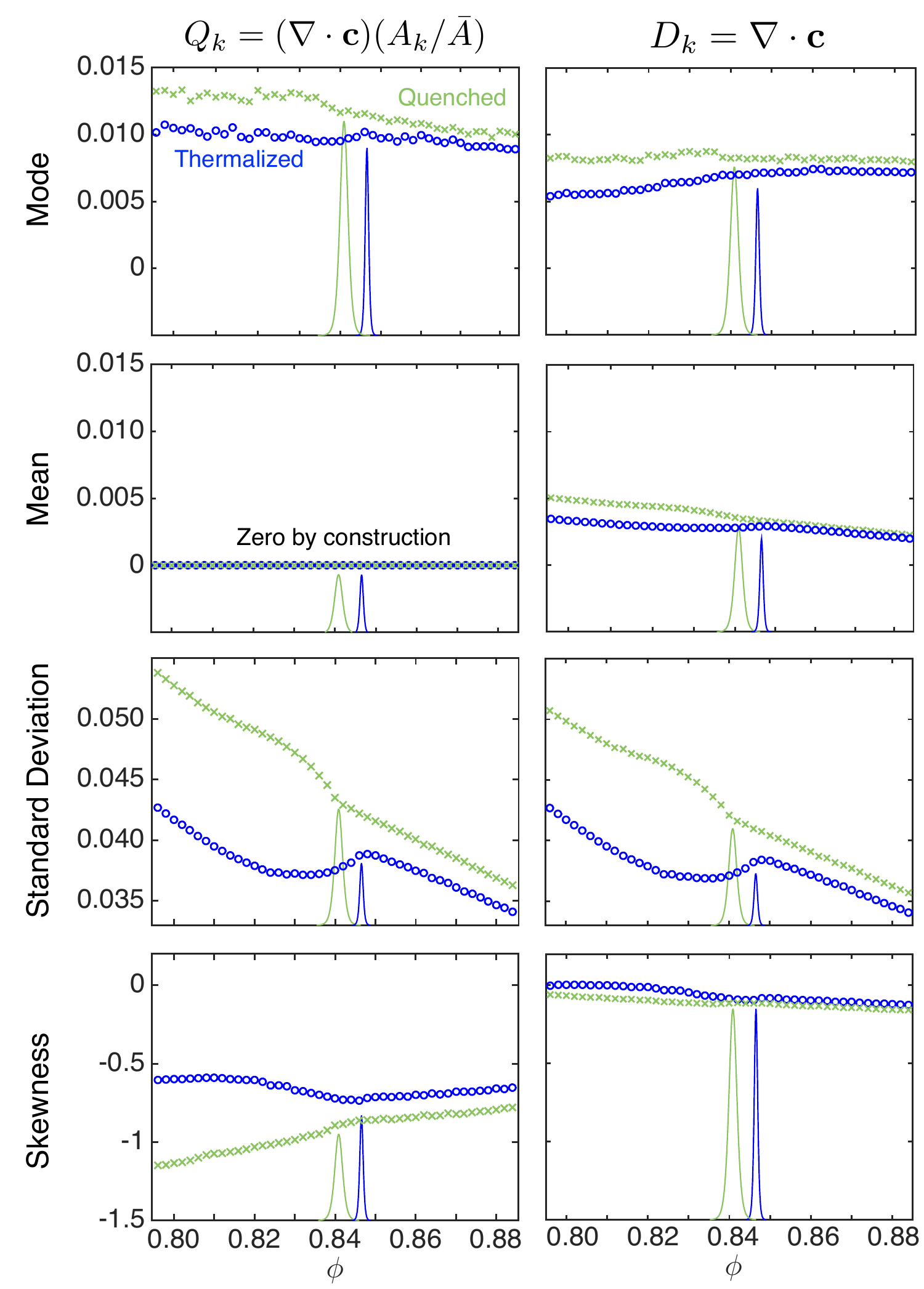}
     \caption{Moments for both area-weighted ($Q_k$, left column) and non-area-wieghted ($D_k$, right column) divergence of the interpolated center-to-centroid vectors.  The modes for both $Q_k$ and $D_k$ are comparable, and there is no noticeable feature in either curve that coincides with the jamming transition.  The mean of $Q_k$ is zero by construction, while the mean of $D_k$ is non-zero.  There is a small feature in the mean of $D_k$ that coincides with each transition.  The standard deviations are very similar, and both have a kink at the transition.  Both skewnesses have kinks at the transition, though the skewnesses are more negative and the kinks are more pronounced in $Q_k$. }
\end{figure*}

%

\begin{figure*} [ht]
     \centering
     \includegraphics[width=1.6\columnwidth]{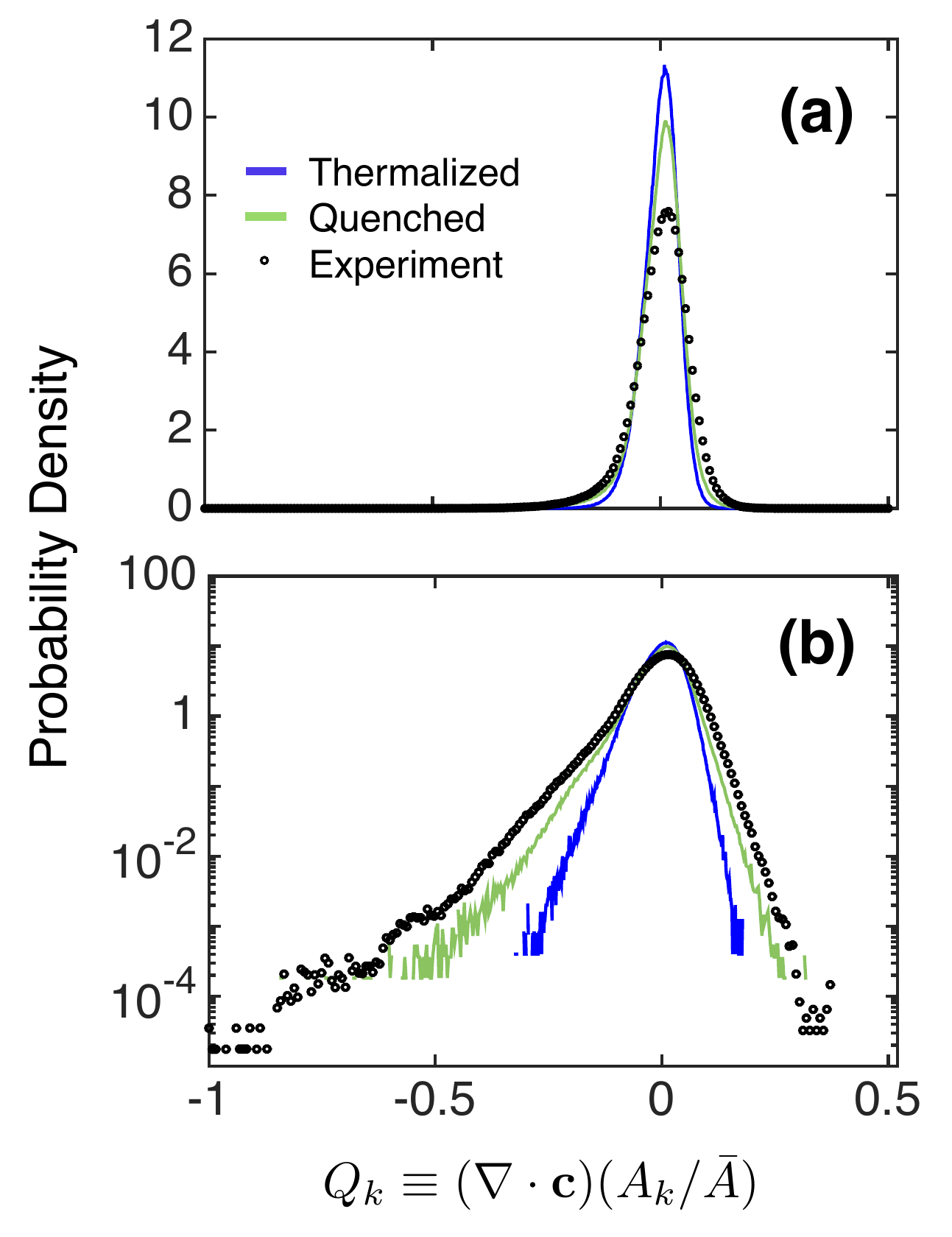}
     \caption{Probability density functions for experimental (black circles), quenched (green), and thermalized (blue) data sets, all of which have packing fraction, $\phi \approx 0.81$.  The linear $y$-axis (a) highlights differences in the peaks of the distributions, and the logarithmic $y$-axis (b) emphasizes the differences in the tails of the distributions.   The experimental data more closely resembles the quenched data, though both numerically-generated packings have sharper peaks and narrower tails than the experiment.  Some of these differences are due to preparation protocol, and some may be due to differing particle size ratios (3:4 in experiment, 1:1.4 in simulation).}
\end{figure*}

\begin{figure*} [ht]
     \centering
     \includegraphics[width=1.6\columnwidth]{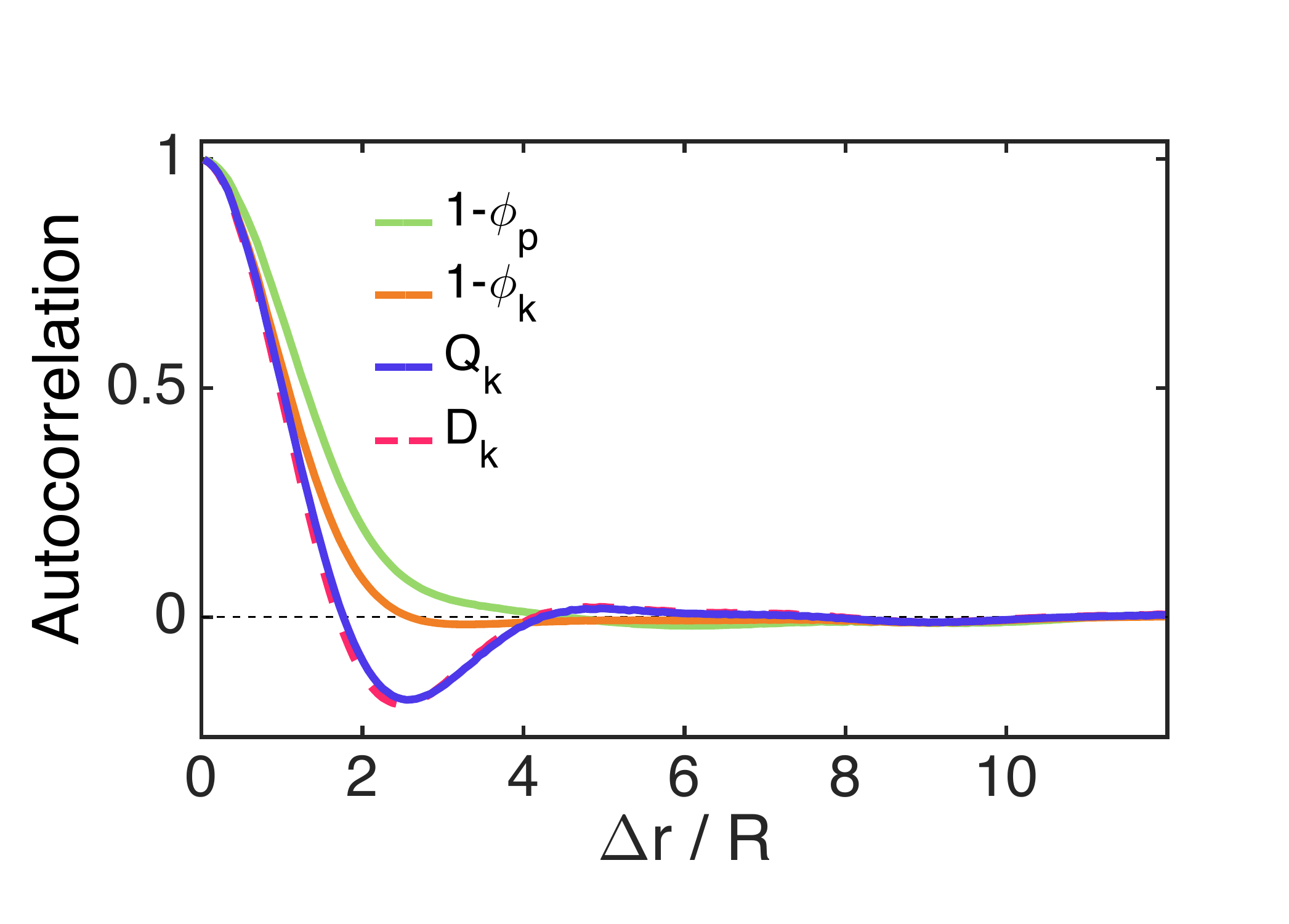}
     \caption{Normalized spatial autocorrelation of four different measures of local free area, at equal times, versus radial distance divided by average particle radius.  These are determined by linearly interpolating each structural measure at every point in time, computing the equal-time two-dimensional autocorrelation of the mean-subtracted interpolations, and normalizing by the value for zero shift. The result is then averaged over angular shifts at each time, and finally averaged over all times.  In all cases, correlations are fairly short-ranged, vanishing by $\Delta r \approx 5$ particle radii.  The negative correlation in both $D_k$ and $Q_k$ is expected as both are measures of free area relative to the immediate surroundings.}
\end{figure*}

\begin{figure*} [ht]
\includegraphics[width=1.6\columnwidth]{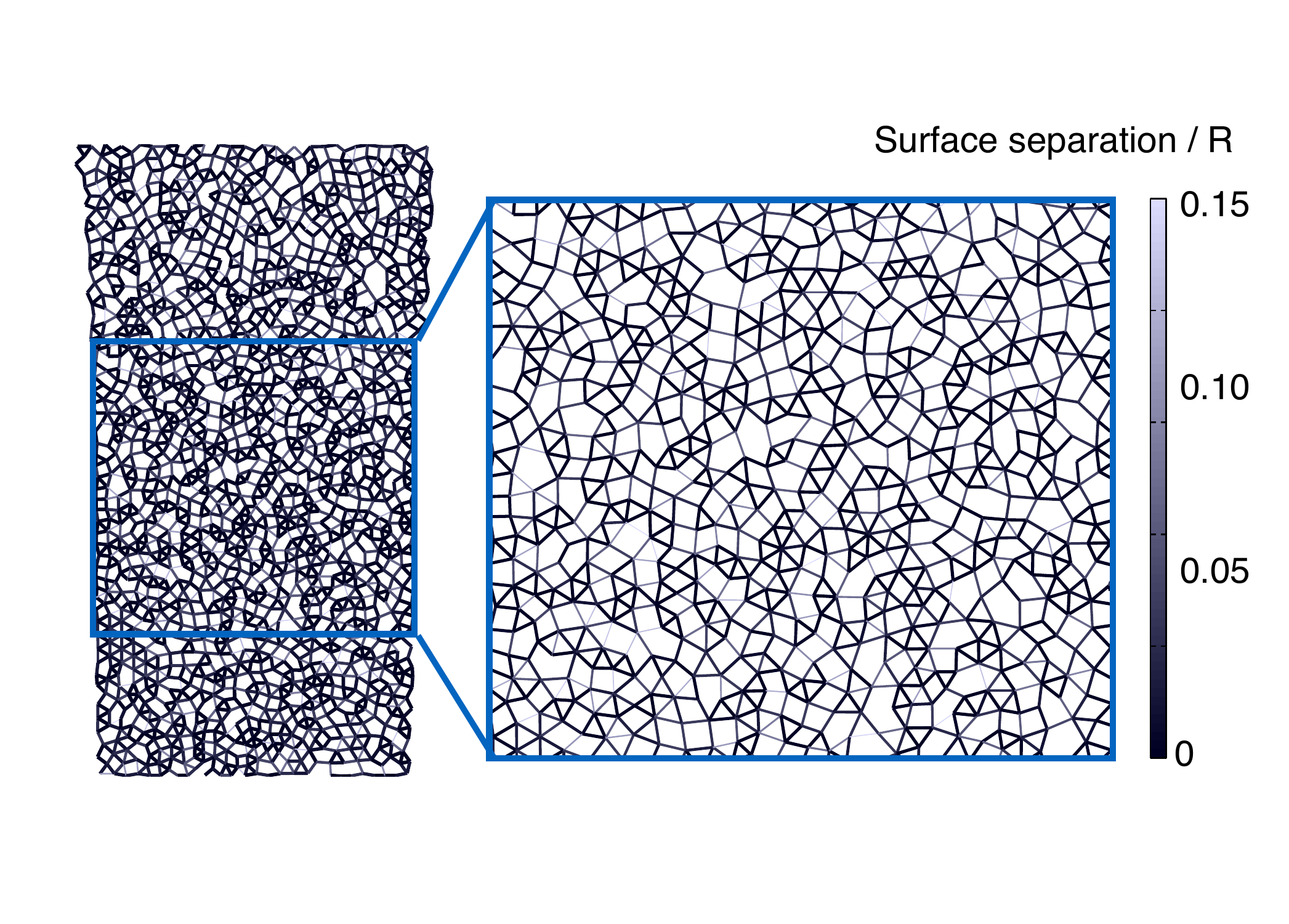}
\caption{Some methods of structural characterization start with a contact or force network. However, this information is not always precisely known, as is the case in for the experimental system shown here.  Each line is drawn from one particle center to an adjacent particle, and the grayscale shading of these links indicates the distance between particle surfaces.  In experimental systems such as this one, there is uncertainty in precise particle locations and size, requiring the choice of a separation threshold to decide which particles are actually in contact.  Lighter colors represent larger separations, so these connections are the first to be pruned if the separation threshold is decreased.  Very small changes in this threshold result in large changes in the resulting network, making this approach unreliable if contact information is imperfectly known as in experiment.  Voronoi-based measures are much more robust to small uncertainties in particle sizes and locations, making them useful in a wider variety of systems.}
\end{figure*}

\end{document}